\documentclass[aps,twocolumn,showkeys]{revtex4}
\usepackage{graphics}
\usepackage[dvips]{graphicx}
\usepackage{times}
\usepackage{multirow}
\usepackage{color}

\begin{document}

\title{The Matthew effect in empirical data}

\author{Matja{\v z} Perc}
\thanks{\href{mailto:matjaz.perc@uni-mb.si}{\textcolor{blue}{matjaz.perc@uni-mb.si}},
\href{http://www.matjazperc.com/}{\textcolor{blue}{matjazperc.com}}}
\affiliation{Faculty of Natural Sciences and Mathematics, University of Maribor, Koro{\v s}ka cesta 160, SI-2000 Maribor, Slovenia}

\begin{abstract}
The Matthew effect describes the phenomenon that in societies the rich tend to get richer and the potent even more powerful. It is closely related to the concept of preferential attachment in network science, where the more connected nodes are destined to acquire many more links in the future than the auxiliary nodes. Cumulative advantage and success-breads-success also both describe the fact that advantage tends to beget further advantage. The concept is behind the many power laws and scaling behaviour in empirical data, and it is at the heart of self-organization across social and natural sciences. Here we review the methodology for measuring preferential attachment in empirical data, as well as the observations of the Matthew effect in patterns of scientific collaboration, socio-technical and biological networks, the propagation of citations, the emergence of scientific progress and impact, career longevity, the evolution of common English words and phrases, as well as in education and brain development. We also discuss whether the Matthew effect is due to chance or optimisation, for example related to homophily in social systems or efficacy in technological systems, and we outline possible directions for future research.
\end{abstract}

\keywords{Matthew effect, preferential attachment, cumulative advantage, self-organization, power law, empirical data}

\maketitle

\section{Introduction}
\label{intro}
The Gospel of St. Matthew states: ``For to all those who have, more will be given.'' Roughly two millennia latter sociologist Robert K. Merton~\cite{merton_sci68} was inspired by this writing and coined ``the Matthew effect'' for explaining discrepancies in recognition received by eminent scientists and unknown researchers for similar work. A few years earlier physicist and information scientist Derek J. de Solla Price~\cite{price_sci65} actually observed the same phenomenon when studying the network of citations between scientific papers, only that he used the phrase cumulative advantage for the description. The concept today is in use to describe the general pattern of self-reinforcing inequality related to economic wealth, political power, prestige, knowledge, or in fact any other scarce or valued resource~\cite{rigney_13}. And it is this type of robust self-organization that goes beyond the particularities of individual systems that frequently gives rise to a power law, where the probability of measuring a particular value of some quantity varies inversely as a power of that value~\cite{zipf_49}. Power laws appear widely in physics, biology, earth and planetary sciences, economics and finance, computer science, demography and the social sciences~\cite{reed_el01, mitzenmacher_im04, newman_cp05, clauset_siam09}. Although there is no single origin of power-law behaviour --- many theories and models have in fact been proposed to explain it~\cite{miller_57, bak_prl87, sneppen_pd97, sornette_pre98, carlson_pre99, gabaix_99, brakman_jrs99, barabasi_s99, jin_pre01, caldarelli_prl02, reed_pre02, pennock_pnas02, vazquez_pre03, reed_pa03, krapivsky_pre05, souza_pnas07, poncela_pone08, baek_njp11, papadopoulos_n12} --- a strong case can be made for the Matthew effect being responsible in many cases. The purpose of this review is to systematically survey research reporting the Matthew effect in empirical data.

\begin{figure}
\centering{\includegraphics[width = 8.5cm]{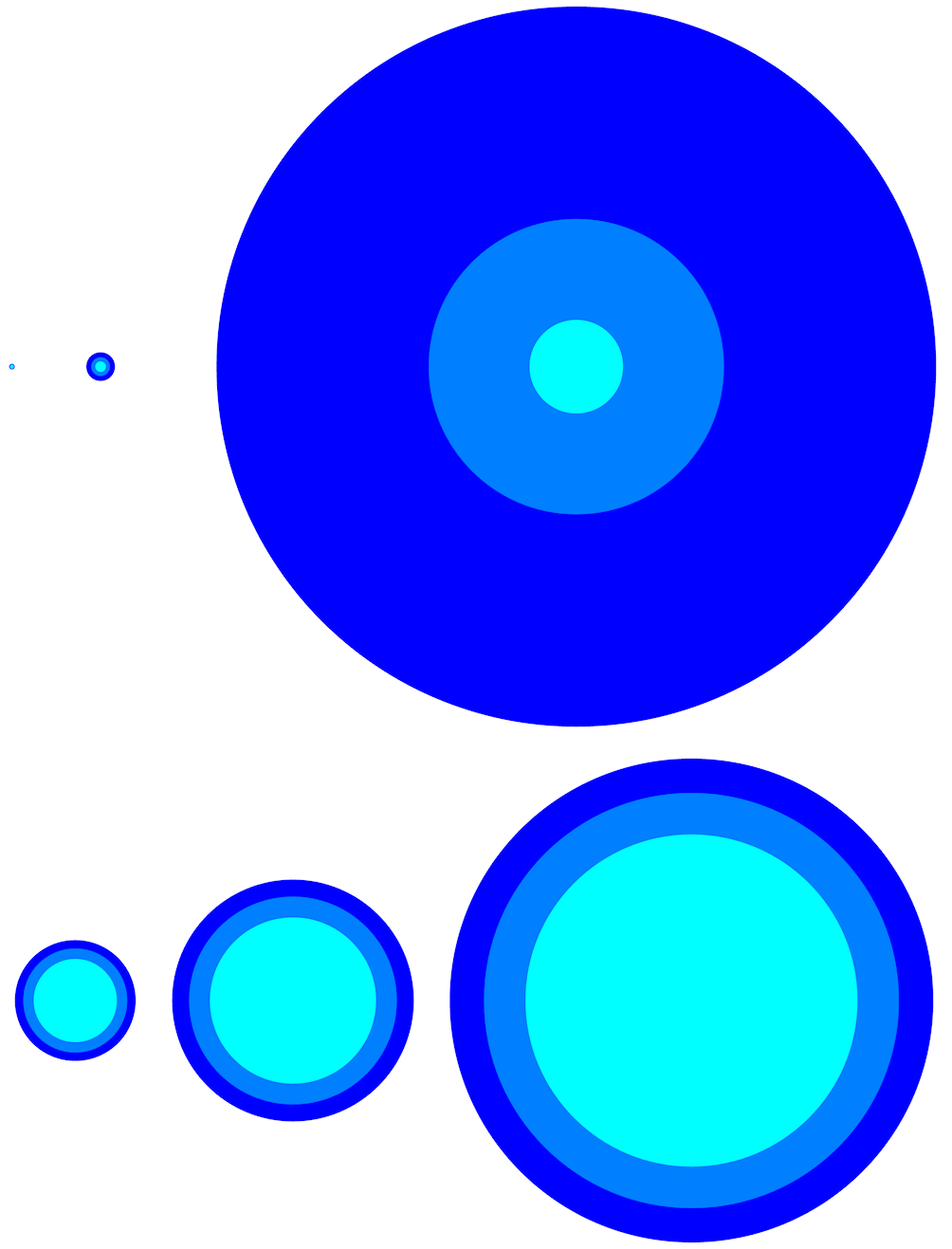}}
\caption{The Matthew effect explained. Top row: Starting with three small circles of practically the same size (small dot on the left), over time, the initial differences grow (middle), until eventually they become massive (right). At the beginning, the blue circle has diameter $5$, light-blue circle has diameter $4$, and the cyan circle has diameter $3$. Assuming the growth is proportional with the size, during each time step the circles may become larger by a factor equivalent to their current diameter. After the first time step (middle), this gives us sizes $25$, $16$ and $9$, respectively. Continuing at the same rate, after the second time step (right), we have sizes $625$, $256$ and $81$. Evidently, such a procedure quickly spirals out of easily imaginable bounds. Bottom row: Taking the logarithm of the same diameters over time (and multiplying by $150$ for visualization purposes only) reveals that, on the log scale, all the circles grow in diameter linearly by a factor of $2$ during each time step from left to right, and the initial relative differences in size remain unchanged over time. This preservation of proportions in logarithmic size manifests as a straight line on a log-log scale --- a power-law distribution. In the depicted schematic example the diameter of the circles can represent anything, from the initial number of collaborators to literacy during formative years.}
\label{scheme}
\end{figure}

In fairness, the Matthew effect has close ties with several other concepts in the social and natural sciences, and it is debatable whether the name we use predominantly throughout this review is the most fitting. The Yule process, inspired by observations of the statistics of biological taxa~\cite{willis_n22}, was in fact the first in a line of widely applicable and closely related mechanisms for generating power laws that relied fundamentally on the assumption that an initially small advantage in numbers may snowball over time~\cite{yule_ptrsb25}. The Gibrat law of proportional growth~\cite{gibrat_31}, inspired by the assumption that the size of an enterprize and its growth are interdependent, also predates the formal introduction of the Matthew effect. Based on the rule of proportional growth, Simon \cite{simon_b55} articulated a stochastic growth model with new entrants to account for the Zipf law~\cite{zipf_49}. The concept of proportional growth has also been elaborated upon thoroughly in Schumpeter's \textit{The Theory of Economic Development} \cite{schumpeter_34}. In terms of popularity and recent impact, however, preferential attachment would without contest be the most apt terminology to use. Barab{\'a}si and Albert~\cite{barabasi_s99} have reasoned that a new node joining a network can in principle connect to any pre-existing node. However, preferential attachment dictates that its choice will not be entirely random, but linearly biased by the number of links that the pre-existing nodes have with other nodes. This induces a rich-get-richer effect, allowing the more connected nodes to gain more links at the expense of their less-connected counterparts. Hence, over time the large-degree nodes turn into hubs and the probability distribution of the degrees across the entire network follows a power law. Although this setup is rather frail as any nonlinearity in the attachment rate may either eliminate the hubs or generate super-hubs~\cite{krapivsky_prl00, dorogovtsev_prl00}, the concept of preferential attachment, along with the ``small-world'' model by Watts and Strogatz~\cite{watts_n98}, undoubtedly helped usher in the era of network science~\cite{watts_99, amaral_pnas00, albert_rmp01, barabasi_02, dorogovtsev_ap02, newman_siamr03, dorogovtsev_03, pastor_04, boccaletti_pr06, newman_06, dorogovtsev_rmp08, barrat_08, arenas_pr08, fortunato_pr10, holme_pr12, barabasi_np12}.

We use the ``Matthew effect'' terminology for practical reasons and to honour the historical account of events, even though the famous writing in the Gospel of St. Matthew might have had significantly different meaning at the time. It was suggested that ``for to all those who have, more will be given'' implied spiritual growth and the development of talents, rather than today's more materialist the ``rich get richer and the poor get poorer'' understanding~\cite{rigney_13}. However, in present times the Matthew effect is appreciated also in education~\cite{stanovich_je08}, so some of the original meaning has apparently been preserved. Whatever the terminology used, the understanding should be that here the Matthew effect stands, at least loosely, for all the aforementioned concepts, including cumulative advantage, proportional growth, and preferential attachment. An illustration of the Matthew effect is presented in Fig.~\ref{scheme}.

Already in their seminal work, Barab{\'a}si and Albert~\cite{barabasi_s99} noted that preferential attachment ought to be readily detected in time-resolved data cataloging network growth. Because of preferential attachment, a node that acquires more connections than another one will increase its connectivity at a higher rate, and thus an initial difference in the connectivity between two nodes will increase further as the network grows, while the degree of individual nodes will grow proportional with the square root of time. This reasoning relates also to the so-called first-mover advantage, which has been found accountable for the remarkable marketing success of certain ahead-of-time products~\cite{kerin_jm92}, as well as the popular acclaim of forefront scientific research despite the fact that it is often less-thorough than follow-up studies~\cite{newman_epl09}. Scientific collaboration networks, where two researchers are connected if they have published a paper together, were among the first empirical data where the concept of preferential attachments has been put to the test and confirmed~\cite{newman_pre01, barabasi_pa02, jeong_epl03, moody_asr04, tomassini_pa07, perc_i10}. Soon to follow were reports of preferential attachment and resulting scaling behaviour in the protein network evolution~\cite{eisenberg_prl03} and the evolution of metabolic networks~\cite{light_bmc05, pfeiffer_pb05}, the Internet \cite{jeong_epl03} and World Wide Web \cite{pennock_pnas02}, the accumulation of citations \cite{jeong_epl03, redner_pt05, valverde_pre07, wang_pa08, wang_pa09, eom_pone11, sheridan_pa12, golosovsky_prl12} and scientific impact \cite{perc_sr13, wang_s13}, the making of new friends and the evolution of socio-technical networks \cite{jeong_epl03, tomkins_KDD06, capocci_pre06, herdaugdelen_epl07, eom_pre08, nazir_acm08, weng_acm13, kunegis_acm13}, population and city size growth \cite{gabaix_99, brakman_jrs99, rozenfeld_pnas08}, the evolution of source code \cite{maillart_prl08} and the most common English words and phrases \cite{perc_jrsi12}, in sexual networks \cite{blasio_pnas07}, as well as the longevity of one's career \cite{petersen_pnas11}, to name but a few examples. Quantitatively less supported but nevertheless plausible arguments in favor of the Matthew effect also come from education, where there is evidence that early deficiencies in literacy may bread lifelong problems in learning new skills \cite{stanovich_je08}, as well as from cognitive neuroscience, where it was hypothesized that the effect could be exploited by means of interventions aimed at improving the brain development of children with low socioeconomic status \cite{raizada_fhn10}. We will review observations of the Matthew effect in empirical data thoroughly in the subsequent Sections, but first we survey the methodology that is commonly employed for measuring preferential attachment.

\section{Measuring preferential attachment}
\label{measure}
The observation of a power law in empirical data \cite{clauset_siam09} might be an indication for the Matthew effect. Importantly, not finding a power-law distribution or at least a related fat-tailed distribution will falsify the Matthew effect, but the opposite does not necessarily hold. Observing a power-law distribution is consistent with the Matthew effect, but indeed many other processes can also generate power-law distributions \cite{mitzenmacher_im04, newman_cp05, sornette_06}. The probability distribution of a quantity $x$ that obeys a power law is
\begin{equation}
p(x) \sim 1 / x^{1+\mu}~~{\rm with}~~\mu >0~,
\label{power}
\end{equation}
where $\alpha=1+\mu$ is the scaling parameter. Since data in the tail ends of power-law distributions is usually very sparse, one has to be careful with the fitting. The usage of maximum-likelihood fitting methods and goodness-of-fit tests based on the Kolmogorov-Smirnov statistics is warmly recommended \cite{clauset_siam09}. Beforehand, there are two ways how to get rid of the noise in the tail, at least visually. One option is to bin the data logarithmically, so that the bins appear evenly spaced on a log scale. The second is to use a cumulative distribution function $q(x) \sim 1 / x^{\mu}$, which gives the probability that the quantity is equal or larger than $x$. In addition to the fact that the later alleviates statistical fluctuations and does not obscure data as do exponentially wider bins, cumulative distributions can also be used to decide on the presence of a power law. Namely, if the probability density function is a power law with the scaling parameter $\alpha$, then the cumulative distribution function should also be a power law, but with an exponent $\alpha-1$. On the other hand, it the probability density function is exponential, the cumulative distribution function will also be exponential, but with the same exponent.

In general, to qualify as a suitable description of empirical data, the probability density function $p(x) \sim 1 / x^{1+\mu}$ should hold within a sufficiently large range of $x$ values, extending over at least two or three decades. It is also advisable that one understands the origin of the deviations from the power law, which often appear at both ends of the distribution. It is also worth pointing out that for $\mu=1$ the power law distribution is commonly referred to as the Zipf law~\cite{zipf_49}, while the cumulative distribution function is the Pareto law \cite{pareto_1895, pareto_97}. The $\mu=1$ case is special because it is at the borderline between the converging and diverging unconditional mean of $x$. While many different physical mechanisms may be at the origin of power laws in complex systems, yielding possibly widely different exponents $\mu$ \cite{mitzenmacher_im04, newman_cp05, sornette_06}, preferential attachment is certainly one viable candidate.

\begin{figure}
\centering{\includegraphics[width = 8.5cm]{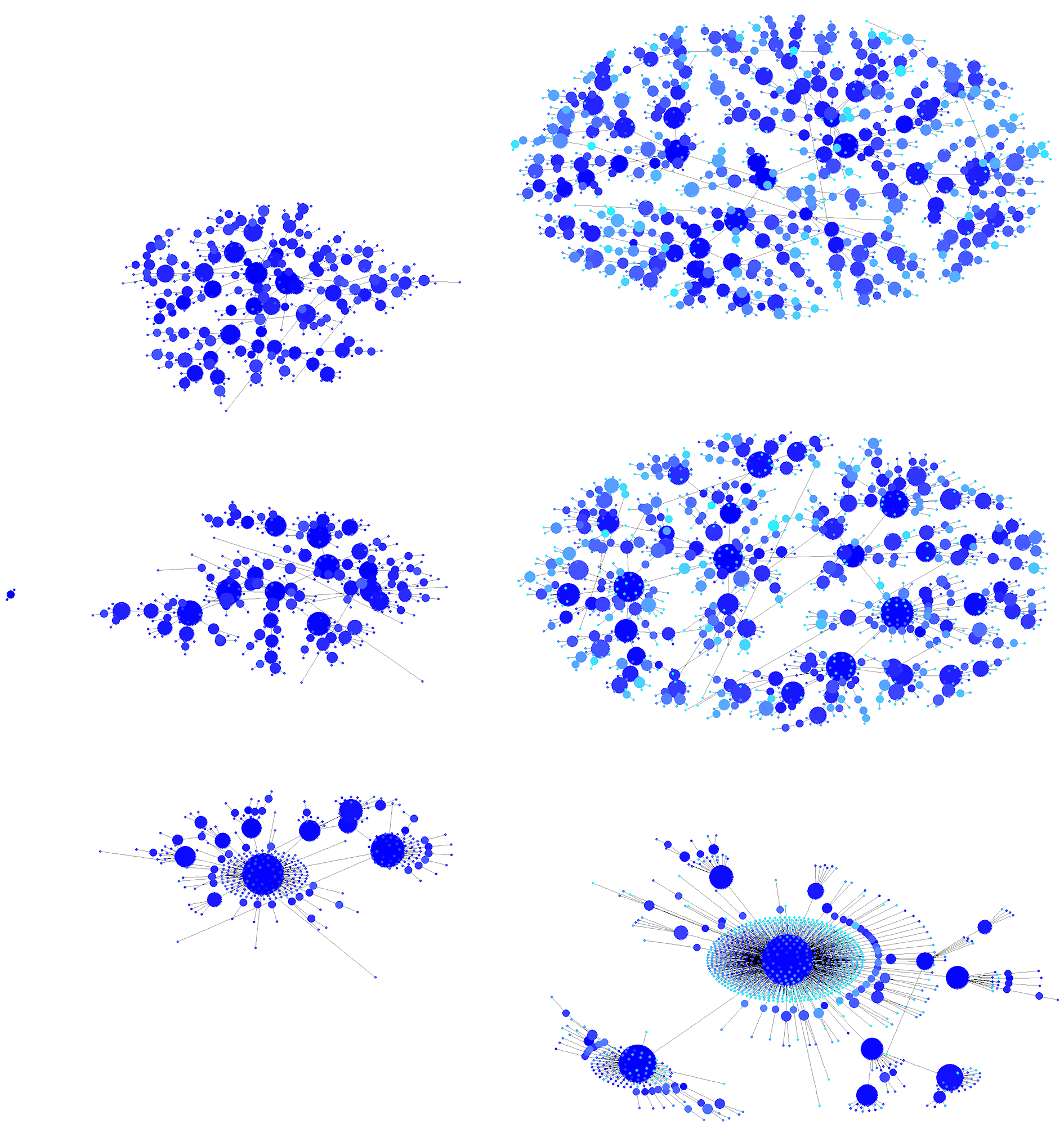}}
\caption{Illustration of network growth by preferential attachment. We start with three nodes, each with a single link to one of the other nodes (small cluster on the left). Subsequently, at each time step, a new nodes arrives and it connects to an existing node with probability $\sim x^\gamma$ (see Eq.~\ref{linear}). Here $x$ is the degree of nodes. After $300$ (center) and $1000$ (right) time steps, sublinear preferential attachment with $\gamma=0.5$ yields the upper two networks, linear preferential attachment with $\gamma=1$ yields the middle two networks, while superlinear preferential attachment with $\gamma=1.5$ yields the lower two networks, respectively. The size and colour (from cyan to blue) of the nodes correspond to their degree in log scale. Sublinear preferential attachment gives rise to a stretched exponential cutoff, thus resulting in somewhat more homogenous networks than linear preferential attachment. Visually, however, the differences are relatively subtle. Superlinear preferential, on the other hand, clearly favours the emergence of ``superhubs'', which attract almost all the nodes forming the network. The complete time evolution of the three networks can be viewed at \href{http://youtu.be/XcGn2KYEmVM}{\textcolor{blue}{youtu.be/XcGn2KYEmVM}},
\href{http://youtu.be/kfuD53o1yKQ}{\textcolor{blue}{youtu.be/kfuD53o1yKQ}} and
\href{http://youtu.be/vB8yI-WrlRg}{\textcolor{blue}{youtu.be/vB8yI-WrlRg}} for $\gamma=0.5$, $\gamma=1$ and $\gamma=1.5$, respectively. Videos for $\gamma=0.25$ and $\gamma=2$, corresponding to even more extreme sublinear and superlinear preferential attachment, are also available at \href{http://youtu.be/85pZodfi4VM}{\textcolor{blue}{youtu.be/85pZodfi4VM}} and
\href{http://youtu.be/85R_AGXk2Ko}{\textcolor{blue}{youtu.be/85R\_AGXk2Ko}}.}
\label{esteban}
\end{figure}

Measuring preferential attachment, however, requires time-resolved data. We need to be able to measure the rate at which all the entities (nodes, papers, people) that make up the studied system acquire the measured quantity $x$ (links, citations, wealth). Assuming the change in $x$ over a short time interval $\Delta t$ is $\Delta x$, the mechanism of preferential attachment assumes that
\begin{equation}
\Delta x \sim A x^\gamma~,
\label{linear}
\end{equation}
where $A$ is the attachment rate and $\gamma$ determines the nonlinearity of the attachment kernel $x^\gamma$. The attachment rate $A$ is time-dependent. In particular, the key assumption underlying the Matthew effect is that $A$ grows proportionally with the growing value of $x$, as schematically depicted in Fig.~\ref{scheme}. However, the preferential attachment mechanism will yield a power law distribution of $x$ values given by Eq.~\ref{power} only if $\gamma=1$, when the attachment kernel is linear \cite{barabasi_s99}. Deviations of $\gamma$ below or above $1$ yield sublinear and superlinear preferential attachment, respectively. Sublinear preferential attachment gives rise to a stretched exponential cutoff, while $\gamma>1$ eventually results in a single entity of the system gaining complete monopoly \cite{krapivsky_prl00, dorogovtsev_prl00}. In the language of growing networks, $\gamma>1$ implies that a single node will over time connect to nearly all other available nodes, while for the accumulation of citations to scientific papers, the superlinear autocatalytic growth may give rise to immortality by means of a dynamical phase transition that leads to the divergence of the citation lifetime of highly cited papers \cite{golosovsky_prl12, golosovsky_jsp13}. The differences, created by different forms of preferential attachment, can be spotted at a glance in the structure of the resulting networks, as shown is Fig.~\ref{esteban}.

A direct application of Eq.~\ref{linear} is problematic because growth governed by preferential attachment is an inherently stochastic process. This statement does not necessarily refer to the origin of preferential attachment --- which is subject to a slowly evolving but very interesting debate on whether the Matthew effect is due to dumb luck or optimisation \cite{barabasi_n12} --- but simply to the fact that, regardless of the origin, there will inevitably be strong irregularities in the way $x$ grows over time for each particular entity of the system. In fact, already Yule's theory of power law distributions in taxonomic groups \cite{yule_ptrsb25} and Champernowne's theory of stochastic recurrence equations \cite{champernowne_e53} showed that there are important links between the Zipf law~\cite{zipf_49} and stochastic growth. More specifically, the autocatalytic growth model actually has the form
\begin{equation}
{\rm d} x \sim \lambda {\rm d} t + \sigma {\rm d} W~,
\label{stoch}
\end{equation}
where $\lambda = A x^\gamma = \frac{\langle \Delta x \rangle}{\Delta t}$ is the average deterministic growth rate over the ensemble of entities with the same $x$ (indicated by $\langle \cdot \rangle$). Moreover, ${\rm d} W$ is an increment of the Wiener process with zero mean and standard deviation $\sigma$. Note also that while $\Delta x$ is a discrete variable and Eq.~\ref{linear} thus essentially a difference equation, $\lambda$ is a continuous variable and Eq.~\ref{stoch} a stochastic differential equation. To do away with the stochastic fingerprint of autocatalytic growth and to estimate reliably whether the process is governed by linear attachment, one can either employ cumulation or averaging. Both methods have been used successfully in the past, although there appear to be persuasive arguments in favor of the latter \cite{golosovsky_jsp13}.

Cumulation was proposed by Jeong et al.~\cite{jeong_epl03}, who used it to test the concept of preferential attachment in a number of different empirical networks. To perform the cumulation, one simply has to calculate
\begin{equation}
\kappa(x)=\int_{0}^{x}\Delta x {\rm d}x~,
\label{cumul}
\end{equation}
where within the integral $x$ is the degree of a node up to a certain time $t$, and $\Delta x$ is the increase in the degree of that same node until $t+\Delta t$. The integration is performed over all the nodes that at time $t$ have degree at most $x$. The sensible expectation is that the stochastic fluctuations in $\Delta x$ will thereby be averaged out, while the key assumption behind the method is that the resulting value of $\kappa(x)$ is the same as if Eq.~\ref{stoch} would be integrated directly over $x$ at a fixed time $t$. Accordingly, we get
\begin{equation}
\kappa(x) \sim A x^{\gamma+1}~,
\label{tada}
\end{equation}
from where one can readily estimate both $A$ and $\gamma$ by fitting $\kappa(x)$ in dependence on $x$. Naturally, we have used the network terminology above only as an example, while of course the same method can be applied on arbitrary time-resolved data to test for preferential attachment \cite{eisenberg_prl03, valverde_pre07, tomassini_pa07, eom_pre08, eom_pone11}.

Averaging, on the other hand, was proposed by Newman~\cite{newman_pre01}, who studied growth and preferential attachment in scientific collaboration networks. In this case, one simply bins the data over $x$, calculates the average growth rate $\lambda = \frac{\langle \Delta x \rangle}{\Delta t}$ for each bin over the ensemble of entities for which $x$ falls within a particular bin (indicated by $\langle \cdot \rangle$), and finally compares the resulting histogram with the prediction of Eq.~\ref{stoch}. The application of this method requires one selects the number of bins to cover the interval of $x$ values, and $\Delta t$ also need not be the finest time-resolution available in the empirical dataset. One can use $\Delta t$ that are larger to further smooth out the fluctuations that might be due to small and intermittent increments of $x$ across short time intervals. In general, it should be possible to select the number of bins and $\Delta t$ such that both $A$ and $\gamma$ could be fitted based on $\lambda = A x^\gamma$ when plotting $\lambda$ in dependence on $x$. This method or a variation thereof has been used in \cite{capocci_pre06, herdaugdelen_epl07, redner_pt05, wang_pa08, perc_i10, perc_jrsi12, perc_sr13}.

While cumulation and averaging are the most frequently applied methods to measure preferential attachment in empirical data, they are not the only ones available. We refer to Golosovsky and Solomon \cite{golosovsky_jsp13} for an in-depth treatment and comparison of the two methods, as well as for an additional control method to check the internal consistency of averaging and cumulation. An additional self-consistent approach to measure preferential attachment in networks has also been proposed in \cite{massen_pa07}, and more recently Markov chain Monte Carlo methodology has been adopted as well \cite{sheridan_pa12}. Interested readers will find further details on how it is possible to improve the measurement of preferential attachment if one is in possession of exceptionally detailed data in \cite{maillart_prl08}, while here we proceed with the review of the Matthew effect in empirical data that stem from an impressive array of different systems.

\section{Scientific collaboration}
\label{collaboration}
We begin with scientific collaboration networks, as they were the first empirical data where the conjectured mechanisms for power-law degree distributions in networks have been put to the test~\cite{newman_pre01, barabasi_pa02, jeong_epl03}. Scientific collaboration networks are a beautiful example of social networks \cite{wasserman_94, watts_99, barabasi_02, christakis_09}, where two researchers are considered connected if they have published a paper together. Notably, for a social network to be representative for what it stands --- an account of human interaction --- a consistent definition of acquaintance is important. And while it may be challenging to define friendship or an enemy in a consistent and precise manner, scientific collaboration is accurately documented in the final product, thus allowing for a precise definition of connectedness and the construction of the social network.

The study of scientific collaboration has been put into the spotlight by the seminal works of Newman~\cite{newman_pre01a, newman_pre01b, newman_pnas01, newman_pnas04}, who constructed networks of connections among researchers by using data from MEDLINE, the Los Alamos e-Print Archive, and NCSTRL. Biomedical research, physics, and computer science were thus comprehensively covered, which helped reveal that some of the discovered structural properties of these networks have a high degree of universality that is beyond scientific disciplines, while other properties of patterns of collaboration, on the other hand, are field-specific. Most notably, it was shown that collaboration networks form ``small worlds'' \cite{newman_pnas01}, in which randomly chosen pairs of researchers are typically separated by only a short path of intermediate acquaintances \cite{watts_n98}. Moreover, the mean and the distribution of the degree of authors revealed the presence of clustering in the networks, which highlighted a number of apparent differences in collaboration patterns between the different fields. The structure of the social science collaboration network has also been studied \cite{moody_asr04}, revealing that a structurally cohesive core in the social sciences has been growing steadily since the early 60s.

\begin{figure}
\centering{\includegraphics[width = 8.5cm]{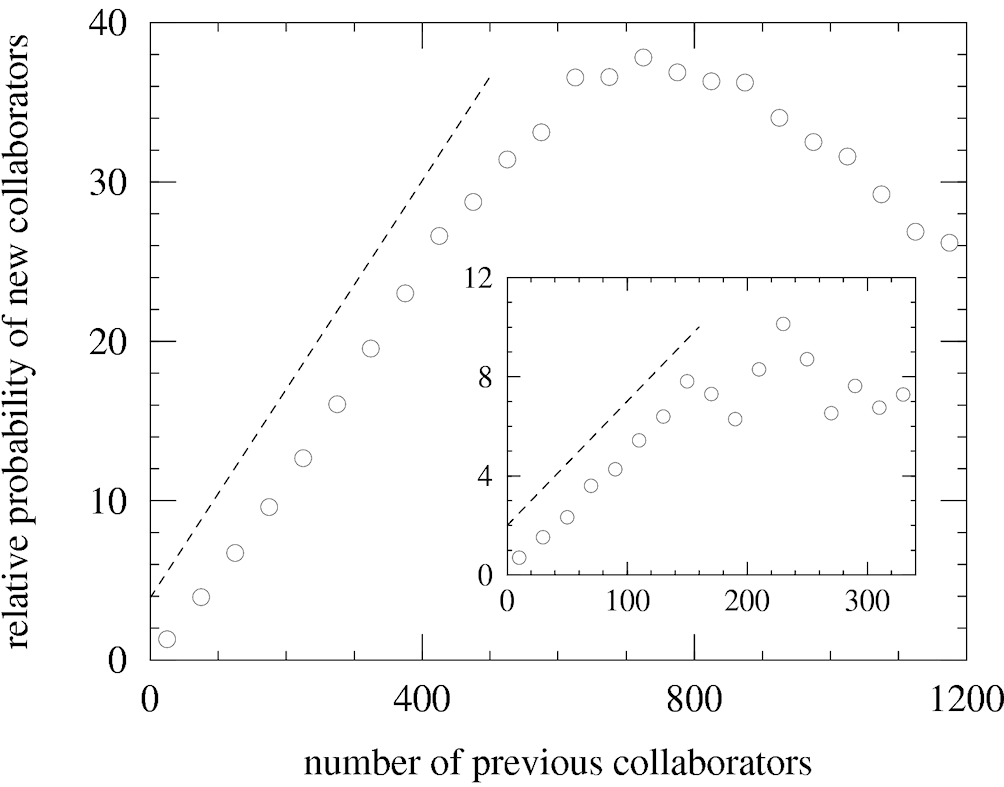}}
\caption{The Matthew effect in scientific collaboration networks. Depicted is the relative probability that a new edge in the network will connect to an author with a given number of previous collaborators. The main panel shows results for the MEDLINE database, while the inset shows results for the Los Alamos e-print Archive. In both cases the relative probability of a new collaborator initially increases linearly with the number of existing collaborators, but there is a field-specific cutoff occurring at around $600$ collaborators in biomedicine (main panel) and $150$ collaborators in physics (inset), which is related to the fundamental limits of scientific collaboration. The figure is reproduced from \cite{newman_pre01} with permission from the American Physical Society.}
\label{pref}
\end{figure}

Practically simultaneously with the research on the structural properties of scientific collaboration networks, research on the time evolution of scientific collaboration networks has been unfolding as well. In~\cite{newman_pre01}, Newman has studied empirically the growth of scientific collaboration networks in physics and biology, employing again data from the Los Alamos e-Print Archive and MEDLINE. It was shown that the probability of a pair of scientists collaborating increases with the number of other collaborators they have in common, and that the probability of a particular scientist acquiring new collaborators increases with the number of his or her past collaborators --- a hallmark property of the Matthew effect. As shown in Fig.~\ref{pref}, which we reproduce from~\cite{newman_pre01}, the relative probability of a new collaborator increases practically linearly with the number of existing collaborators. This is particularly true for the initial part of the curve, but since no one can collaborate with an infinite number of people in a finite period of time, the probability falls off as $x$ (here denoting the degree of authors) becomes large. Interestingly, this point appears to be around $150$ collaborators in physics (inset) and $600$ in biomedicine (main panel), indicating the aforementioned differences in the patterns of collaboration between scientific disciplines.

A closer look at the results presented in Fig.~\ref{pref} reveals that the employed averaging method actually yields $\gamma=1.04$ for MEDLINE and $\gamma=0.89$ for the Los Alamos e-Print Archive, which in agreement with Eq.~\ref{linear} corresponds to slightly superlinear and sublinear preferential attachment, respectively. A closely related study that was conducted around the same time by Barab{\'a}si et al.~\cite{barabasi_pa02}, and which was based on all relevant journals in mathematics and neuroscience, also produced evidence for sublinear preferential attachment with $\gamma=0.8$. The growth of Slovenia's scientific collaboration network~\cite{perc_i10} and a coauthorship study based on neuroscience journals~\cite{jeong_epl03} also supported the concept of sublinear preferential attachment, both reporting $\gamma=0.79$. The lowest $\gamma$ value was reported by Tomassini et al.~\cite{tomassini_pa07}, who showed that the time evolution of the genetic programming coauthorship network is governed by $\gamma=0.76$. However, time-reversing or permutating randomly the order in which the coauthorship networks were constructed within the resolution window $\Delta t$ yielded $\gamma=0.88$ and $\gamma=0.85$, respectively. Taken together, these results favour the concept of slightly sublinear preferential attachment governing the growth of scientific collaboration networks, but as rightfully pointed out by Newman~\cite{newman_pre01}, in alternative to linear preferential attachment this difference may have little effect. As shown by Krapivsky et al.~\cite{krapivsky_prl00} and reviewed in Section~\ref{measure}, sublinear preferential attachment gives rise to a stretched exponential cutoff in the resulting degree distribution, but a similar cutoff is already present in the degree distribution as a result of the deviation from linear behaviour for sufficiently large $x$ in Fig.~\ref{pref}. Indeed, the same deviation has also been reported for the growth of Slovenia's scientific collaboration network~\cite{perc_i10}, thus providing evidence that the sublinear preferential attachment translates fairly accurately into the expected degree distribution.

Irrespective of these details, the overwhelming evidence fully supports the Matthew effect in scientific collaboration networks, indicating that over time initial differences in the number of collaborators are destined to grow and give rise to a strong segregation among authors. Ultimately, some individuals therefore acquire hundreds while others only a handful of collaborators during their scientific career.

\section{Socio-technical and biological networks}
\label{networks}
Scientific collaboration networks reviewed above are obviously also prime examples of social networks and would thus be fit for this section, but we have awarded them a separate section due to their forerunner role in testing preferential attachment in empirical data. There are, however, a number of other socio-technical \cite{jeong_epl03, tomkins_KDD06, capocci_pre06, herdaugdelen_epl07, eom_pre08, maillart_prl08, rozenfeld_pnas08, weng_acm13, kunegis_acm13} and biological \cite{eisenberg_prl03, pfeiffer_pb05} networks, where the availability of time-resolved data allowed testing for the Matthew effect. The evolution of socio-technical networks in particular has been in the focus of attention for decades \cite{trist_e81}. Recent leaps of progress in the availability of reliable ``big data'', mathematical modelling and informatics tools enable increasingly deeper understanding of contagion processes, emerging tipping points, cascading and related nonlinear phenomena that underpin the most interesting characteristics of socio-technical systems \cite{vespignani_np12, borge_nets13}.

The Matthew effect in socio-technical networks has been reported first by Jeong et al.~\cite{jeong_epl03}, who at that time also proposed cumulation (see Eqs.~\ref{cumul} and \ref{tada}) to measure preferential attachment in time-resolved data describing network growth. In addition to a scientific collaboration network (see Section~\ref{collaboration}) and a citation network (see Section~\ref{citations}), they have shown that the evolution of the network of movie actors and the evolution of the autonomous systems forming the Internet are both governed by near-linear preferential attachment. Akin to the definition of a scientific collaboration network, in the movie actor network two actors are connected if they have acted together in a movie. The investigated network was made up of all movies and actors from $1892$ till $1999$, and it was shown that the growth is characterized by $\gamma=0.81$. Similarly as by scientific collaboration, here too the slightly sublinear character of preferential attachment can be linked to obvious constrains in the number of co-actors an individual can possibly amass in the course of a lifetime, and this also translates to the expected exponential cutoff in the resulting degree distribution of actors. Notably, preferential attachment in a movie actor network was also reported in \cite{eom_pre08}. For the Internet, Jeong et al.~\cite{jeong_epl03} used the data provided by NLANR, and they have observed slightly superlinear preferential attachment characterized by $\gamma=1.05$. As evidenced by the examples of network growth depicted in Fig.~\ref{esteban}, however, such small deviations from $\gamma=1$ lead to hardly recognizable deviations (note that in the depicted examples we have used $\gamma=0.5$ for sublinear and $\gamma=1.5$ for superlinear preferential attachment), and one can thus in good faith conclude to the Matthew effect as a more general description of the mechanism governing the growth of these networks.

In addition to the Internet, the related World Wide Web has also been shown to displays striking rich-get-richer behaviour that is driven by the competition of links on the web \cite{pennock_pnas02, herdaugdelen_epl07}. Interestingly, although the connectivity distribution over the entire web is close to a pure power law, Pennock et al.~\cite{pennock_pnas02} reported that the distribution within sets of category-specific web pages is typically unimodal on a log scale, with the location of the mode, and thus the extent of the rich get richer phenomenon, varying across different categories. A simple generative model, incorporating a mixture of preferential and uniform attachment to describe these observation has also been proposed \cite{pennock_pnas02}.

\begin{figure}
\centering{\includegraphics[width = 8.5cm]{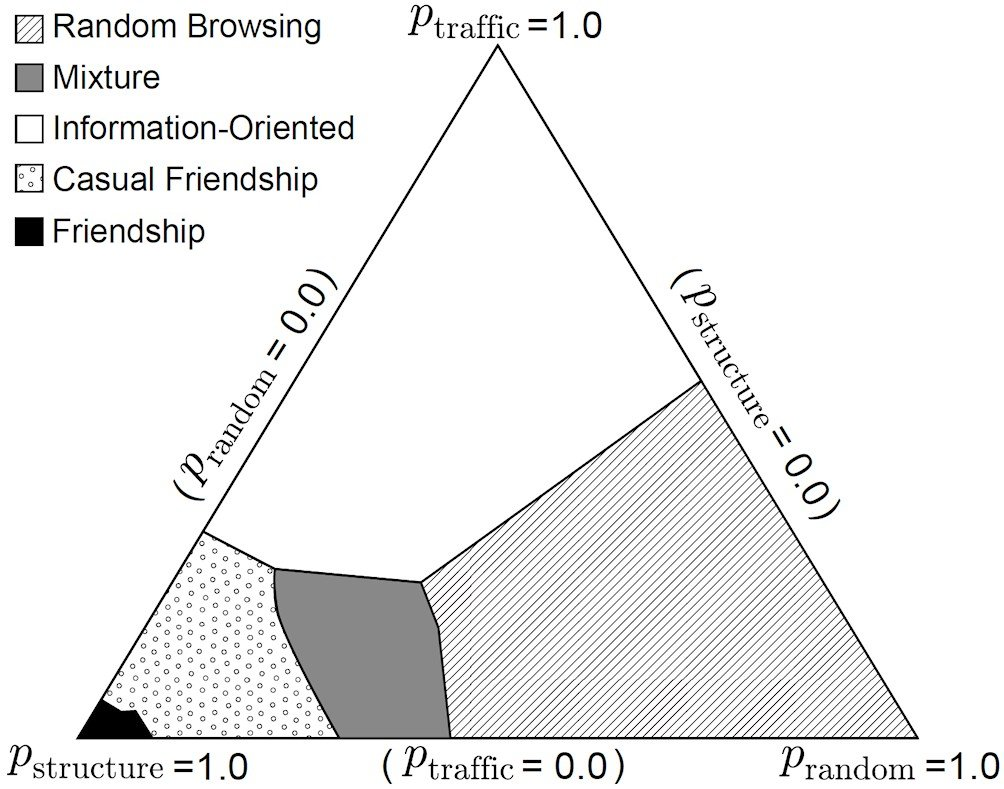}}
\caption{The expansion of online social circles is governed by users that employ many different individual link creation strategies. Indeed, various criteria are taken into account in different proportions when deciding with whom to connect next. The depicted ternary plot encodes the proportions of different link creation strategies for different user types (see legend) in terms of structure ($p_{{\rm structure}}$), traffic ($p_{{\rm traffic}}$) and chance ($p_{{\rm random}}$). Combined, these strategies may give rise to the Matthew effect and lead to strongly heterogeneous social interaction networks. The figure is reproduced from \cite{weng_acm13} with permission from the ACM.}
\label{users}
\end{figure}

Online social networks, such as the internet encyclopedia Wikipedia \cite{capocci_pre06}, bulletin board systems \cite{eom_pre08}, social networking services like Flickr, the obsolete Yahoo! $360^\circ$ or the now popular Facebook \cite{tomkins_KDD06, nazir_acm08}, as well as longitudinal micro-blogging data \cite{weng_acm13} also show evidence of the Matthew effect. Wikipedia growth, for example, can be described by local rules such as the preferential attachment
mechanism, despite the fact that individual users who are responsible for its evolution can act globally on the network \cite{capocci_pre06}. Research also revealed that triadic closure --- if Alice follows Bob and Bob follows Charlie, Alice will follow Charlie --- is not such a major mechanism for creating social links in online networks as initially assumed. Longitudinal micro-blogging data reveal more complex strategies that are employed by users when expanding their social circles \cite{weng_acm13}. In particular, while the network structure affects the spread of information among users, the network is in turn shaped by this communication activity. This suggests a link creation mechanism whereby Alice is more likely to follow Charlie after seeing many messages by Charlie. Weng et al.~\cite{weng_acm13} conclude that triadic closure does have a strong effect on link formation, but shortcuts based on traffic are another key factor in interpreting network evolution. Link creation behaviours can be summarized by classifying users in different categories with distinct structural and behavioural characteristics, as shown in Fig.~\ref{users}. Users who are popular, active, and influential tend to create traffic-based shortcuts, making the information diffusion process more efficient in the network \cite{weng_acm13}. Notably, the subject of preferential attachment in online networks has recently been surveyed comprehensively in \cite{kunegis_acm13}, where interested readers will find many further examples and interesting information related specifically to this type of empirical data.

In addition to the vast landscape of online social networks, there are also many socio-technical systems that do not exist solely online, but for which useful data can still be obtained. Rozenfeld et al.~\cite{rozenfeld_pnas08}, for example, introduced a method to designate metropolitan areas called the ``City Clustering Algorithm'', and used the obtained data to examine the Gibrat law of proportional growth~\cite{gibrat_31}. The latter postulates that the mean and standard deviation of the growth rate of cities are constant, independent of city size. The study revealed that the data deviate from the Gibrat law, and that the standard deviation decreases as a power law with respect to the city size. The ``City Clustering Algorithm'' allowed for the study of the underlying process leading to these deviations, which were shown to arise from the existence of long-range spatial correlations in population growth. Prior to this empirical research, Gabaix~\cite{gabaix_99} and Brakman et al.~\cite{brakman_jrs99} elaborated theoretically on the mechanisms behind city growth, including prominently on the Zipf law.

Maillart et al.~\cite{maillart_prl08}, on the other hand, made use of detailed data on the evolution of open source software projects in Linux distributions. They have showed that the network resulting from the tens of thousands of connected packages precisely obeys the Zipf law over four orders of magnitude, and that this is due to stochastic proportional growth. The study thus delivers a remarkable example of a growing complex self-organizing adaptive system that is subject to the Matthew effect.

Sexual contact networks have also been the subject of research related to the Matthew effect \cite{jones_procb03, blasio_pnas07}. In particular, de Blasio et al.~\cite{blasio_pnas07} have tested the conjecture of preferential attachment by means of a maximum likelihood
estimation-based expectation-maximization fitting technique, which was used to model new partners over a 1-year period based on the number of partners in foregoing periods of two and four years, as well as the lifetime. The preferential attachment model was modified to account for individual
heterogeneity in the inclination to find new partners and fitted to Norwegian survey data on heterosexual men and women. The research revealed sublinear preferential attachment governing the growth of sexual contact networks with $0.5 \leq \gamma \leq 0.7$, which similarly like for scientific collaboration and movie actor networks reviewed above, likely has to do with the physical limits of sexual contacts. Interestingly, the lower value of $\gamma$ might suggest that the constrains on the maximal feasible number of sexual partners are greater than on the number of collaborators or co-actors in a movie, thus leading to a stronger exponential cutoff in the corresponding probability distributions  --- a conclusion that certainly seems to resonate with reality. Moreover, a preceding study by Jones and Handcock~\cite{jones_procb03} concluded that the scaling of sexual degree distributions and the underlying assumption of preferential attachment is actually a very poor fit to the data stemming from several different sexual contact networks. This in turn has important implications for reducing the transmissibility of sexually transmitted diseases, for example by means of condom use or high-activity anti-retroviral therapy, as such interventions could thus bring a population below the epidemic transition, even in populations exhibiting large degrees of behavioural heterogeneity.

To conclude this Section, we review examples of the Matthew effect in biological networks, where in relation to the socio-technical networks, the examples are comparatively few. The \textit{Saccharomyces cerevisiae} protein-protein interaction network \cite{mering_n02} has a scale-free topology, and Eisenberg and Levanon~\cite{eisenberg_prl03} have shown that the older a protein the better connected it is, and that the number of interactions a protein gains during its evolution is proportional to its connectivity. Thus, by using a cross-genome comparison, the study shows conclusively that the evolution of protein networks is governed by linear preferential attachment. Eisenberg and Levanon go on to conclude that preferential attachment is an important concept in the process of evolution, as it dynamically leads to the formation of big protein complexes and pathways, which introduce high complexity regulation and functionality~\cite{eisenberg_prl03}.

The Matthew effect has also been studied in metabolic networks~\cite{light_bmc05, pfeiffer_pb05}, which are at the heart of interactions between biochemical compounds in living cells. Light et al.~\cite{light_bmc05} have determined the connectivity patterns of enzymes in the metabolic network of \textit{Escherichia coli}, showing that enzymes which have representatives in eukaryotes have a higher average degree, while enzymes which are represented only in the prokaryotes, and especially the enzymes only present in $\beta \gamma$-proteobacteria, have a lower degree than expected by chance. More importantly, the research revealed that new edges are added to the highly connected enzymes at a faster rate than to the enzymes with low degree, which is consistent with the Matthew effect. The proposed biological explanation for the observed preferential attachment in the growth of metabolic networks was that novel enzymes created through gene duplication maintain some of the compounds involved in the original reaction throughout its future evolution. Although it remains a major challenge in biology to understand the causes and consequences of the specific design of metabolic networks, Pfeiffer et al.~\cite{pfeiffer_pb05} have shown that the reported empirical observations, in particular the characteristic presence of hub metabolites such as ATP or NADH, could be explained by computer simulations that initially involve only a few multifunctional enzymes. Then, through the selection of growth rates governed by essential biochemical mechanisms, hubs emerge spontaneously through the process of enzyme duplication and specialization.

\section{Citations}
\label{citations}
After the rather extensive but hopefully interesting departure from scientific collaboration networks to socio-technical and biological networks, we may refocus on research, in particular on the accumulation of citations to scientific papers. Researchers seem to delight in meticulously evaluating their scientific output and its impact. From citation distributions \cite{seglen_92, redner_epjb98, lehmann_pre03, radicchi_pnas08, stringer_amsoc10, radicchi_pone12}, coauthorship networks \cite{newman_pnas04} and the formation of research teams \cite{guimera_s05, milojevic_pnas14}, to the ranking of researchers \cite{hirsch_pnas05, radicchi_pre09, petersen_pre10} and the predictability of their success \cite{acuna_n12, penner_sr13, penner_pt13, wang_s13} --- how we do science has become a science in its own right. Not surprisingly, the patterns of citation accumulation have been, just like the evolution and structure of scientific collaboration networks, studied extensively during the past decade \cite{jeong_epl03, redner_pt05, valverde_pre07, wang_pa08, wang_pa09, eom_pone11, golosovsky_prl12, golosovsky_jsp13}.

Notwithstanding the seminal observations by Robert K. Merton~\cite{merton_sci68}, who actually introduced the Matthew effect based on the discrepancies in recognition received by eminent scientists and unknown researchers for similar discoveries, and the work by Derek J. de Solla Price~\cite{price_sci65}, who was studying the network of citations between scientific papers already in the early 60s, the first more rigorous test of preferential attachment in the accumulation of citations is again due to Jeong et al.~\cite{jeong_epl03}. They have shown that the citations to papers published in the \textit{Physical Review Letters} since $1989$ accumulate by means of slightly sublinear preferential attachment with $\gamma=0.95$. Soon thereafter, Redner~\cite{redner_pt05} conducted an analysis of the entire citation history of publications of \textit{Physical Review}, at the time spanning $110$ years, and also confirmed that linear preferential attachment appears to account for the propagation of citations. At closer inspection, the analysis even hinted towards slightly superlinear accumulation, although this, as well as the prospect of strictly linear preferential attachment, was in disagreement with the reported log-normal distribution of citations. Two papers by Wang et al.~\cite{wang_pa08, wang_pa09}, using as empirical data citations to papers published in the \textit{Journal of Applied Physics} between $1931$ and $2005$, the \textit{Journal of Experimental Medicine} between $1900$ and $2005$, and the \textit{IEEE Transactions on Automatic Control} between $1963$ and $2005$, delivered essentially the same results, reporting $\gamma \approx 1$ to govern the accumulation of citations. Eom and Fortunato \cite{eom_pone11} also used the full publication history of the \textit{Physical Review} minus \textit{Reviews of Modern Physics} to study the evolution of citation networks, and they have proposed a linear preferential attachment model with time dependent initial attractiveness that successfully reproduces the empirical citation distributions as well as accounts for the presence of observed citation bursts.

Importantly, the accumulation of citations to scientific papers has recently been revisited by Golosovsky and Solomon~\cite{golosovsky_prl12}, who confirmed the hints reported already by Redner \cite{redner_pt05}, namely that the citation dynamics is nevertheless governed by superlinear preferential attachment with $1.25 \leq \gamma \leq 1.3$. The research used as data the citation history of $40195$ physics papers published in one year, and it was emphasized that the citation process cannot be described as a memoryless Markov chain since there is a substantial correlation between the present and recent citation rates to a paper. Based on these observations, a stochastic dynamical model of a growing citation network based on a self-exciting point process has been proposed, and it was demonstrated that it accounts perfectly for the measured citation distributions. An intriguing consequence of this result is that the superlinear autocatalytic growth conveys immortality to highly cited papers by means of a dynamical phase transition that leads to the divergence of the citation lifetime --- in the language of epidemiology, these papers become endemic \cite{golosovsky_prl12, golosovsky_jsp13}.

Lending further support to the conclusions of Golosovsky and Solomon~\cite{golosovsky_prl12} are several preceding accounts of superlinear preferential attachment in the accumulation of citations, however not to scientific papers, but rather to patents \cite{valverde_pre07, sheridan_pa12}. Valverde et al~\cite{valverde_pre07}, for example, studied the patent citation network resulting from the patents registered by the U.S. Patent and Trademark Office, and in the light of similarities with article citation networks, concluded towards a universal type of mechanism that links ideas, designs as well as their evolution. This mechanism can be broadly classified as the Matthew effect, which governs how credit is amassed by research as well as technological innovations.

Notably, the subject of preferential attachment in the accumulation of citations has recently been surveyed comprehensively in \cite{golosovsky_jsp13}, where interested readers will find further interesting information related specifically to this type of empirical data.

\section{Scientific progress and impact}
\label{impact}

\begin{figure}
\centering{\includegraphics[width = 8.5cm]{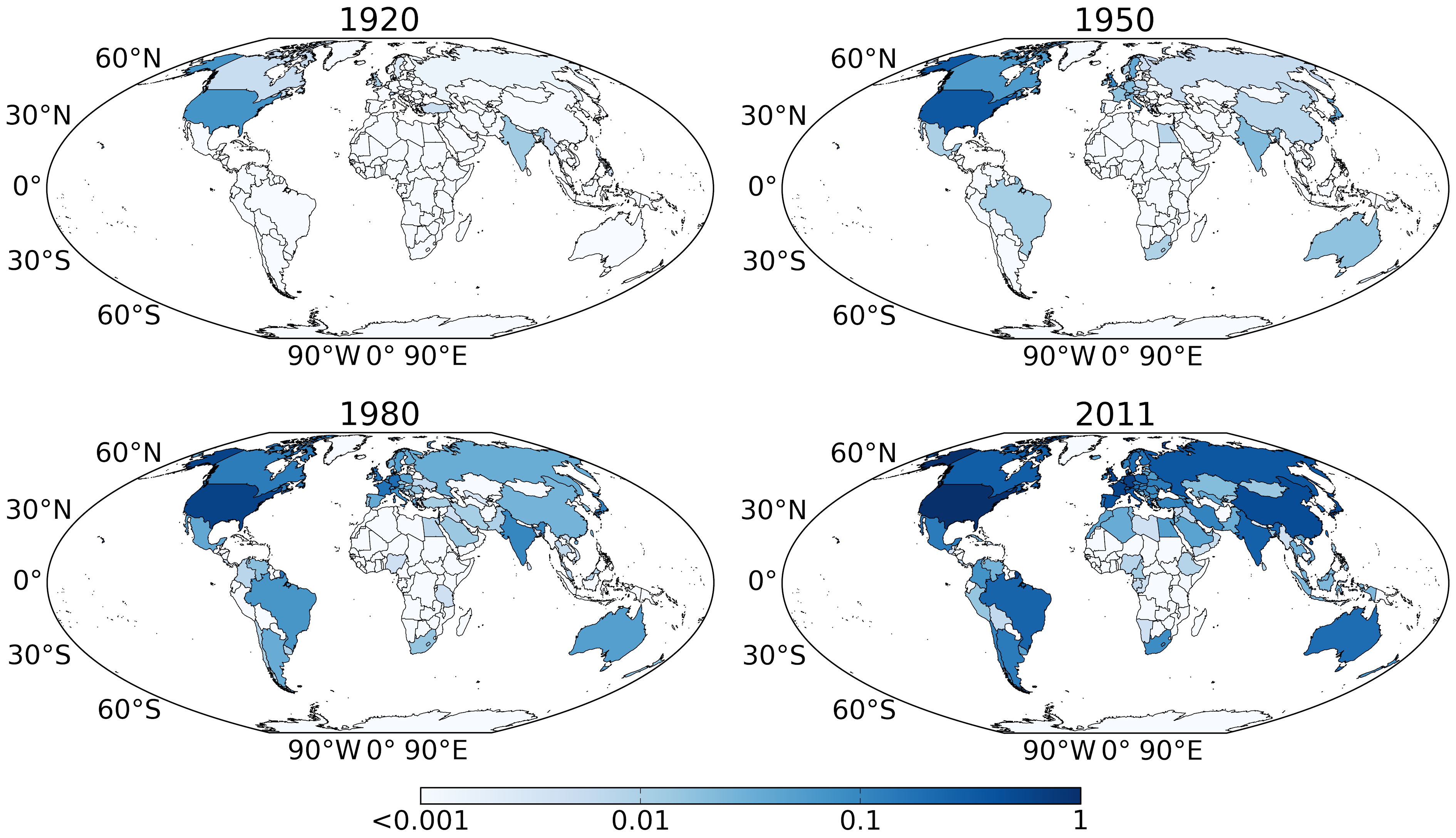}}
\caption{Countries that contribute to research that is published in the Physical Review. Colour encodes the average monthly productivity of a country during each displayed year, normalized by the average monthly output of the U.S. during 2011 (equalling $\approx 565$ publications per month --- a maximum). All affiliations were used, and in case more than one country was involved on a given publication, all received equal credit. A 12 month moving average was applied prior to calculating the average monthly production for each country. Note that the colour scale is logarithmic. Displayed are World maps for four representative years, while the full geographical timeline can be viewed at \href{http://youtu.be/0Xeysi-EfZs}{\textcolor{blue}{youtu.be/0Xeysi-EfZs}}. The figure is reproduced from \cite{perc_sr13}.}
\label{maps}
\end{figure}

The Matthew effect in the evolution of scientific collaboration networks and in the propagation of citations begets the question whether scientific progress and impact in general might be subject to the same effect. The increasing availability of vast amounts of digitised data, in particular massive databases of scanned books \cite{michel_s11} as well as electronic publication and informatics archives \cite{evans_s11}, fuel large-scale explorations of the human culture that were unimaginable even a decade ago. And since science is central to many key pillars of the human culture, the science of science is scaling up massively as well, with studies on World citation and collaboration networks \cite{pan_sr12}, the global analysis of the ``scientific food web'' \cite{mazloumian_scirep13}, and the identification of phylomemetic patterns in science evolution \cite{chavalarias_pone13}, culminating in the visually compelling atlases of science \cite{borner_10} and knowledge \cite{borner_14}.

Riding on the wave of increasing availability of digitised data is also the study of scientific impact, which is gaining on momentum rapidly~\cite{radicchi_scirep12, penner_sr13, penner_pt13, uzzi_s13, wang_s13}. Recent research has revealed, for example, that there is ``no bad publicity'' in science since criticized papers are in fact highly impactful~\cite{radicchi_scirep12}, and that atypical combinations in science have a higher chance to make a big impact~\cite{uzzi_s13}. Clear limits have also been established on the predictability of future impact in science~\cite{penner_sr13, penner_pt13}, contrary to the overly optimistic predictions reported earlier \cite{acuna_n12}. Wang et al.~\cite{wang_s13} have recently proposed a mechanistic model for the quantification of long-term scientific impact, which allows to collapse the citation histories of papers from different journals and disciplines into a single curve, indicating that all papers tend to follow the same universal temporal pattern. The study revealed that the proposed lognormal model without preferential attachment is able to correctly capture only the citation history of small impact papers, while the modelling of the citation patterns of medium and high impact papers requires preferential attachment be turned on. In fact, the model has enabled the team to make an analytical prediction of the citation threshold when preferential attachment becomes relevant, which was reported to equal $8.5$~\cite{wang_s13}. Hence, the impact of papers that surpass this threshold will benefit from the Matthew effect, while papers with fewer citations will not. Wang et al.~\cite{wang_s13} also emphasized that the reported analytical prediction is in close agreement with the empirical finding that preferential attachment is masked by initial attractiveness for papers with fewer than seven citations, as reported earlier by Eom and Fortunato~\cite{eom_pone11}.

\begin{figure}
\centering{\includegraphics[width = 7.5cm]{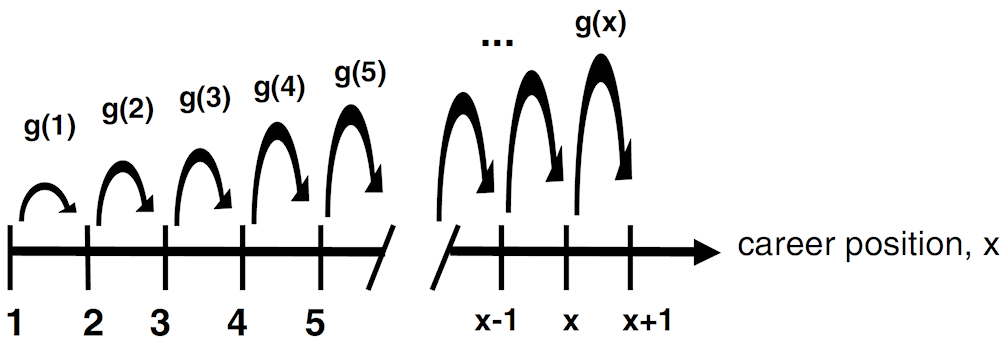}}
\caption{The Matthew effect in professional careers. Progress from career position $x$ to career position $x+1$ is made with a position-dependent progress rate $g(x)=1-\exp[-(x/x_c)^\gamma]$, which increases from approximately zero and asymptotically approaches one over a characteristic time interval $x_c$. For $x \ll x_c$ the progress rate corresponds to $g(x)\sim x^\gamma$, which for $\gamma=1$ is the traditional ansatz for linear preferential attachment (see Eq.~\ref{linear}). Since the progress rate increases with increasing $x$, the essence of the Matthew effect is taken into account in that it becomes easier to make progress the further along the career an individual is. The figure is reproduced from \cite{petersen_pnas11}.}
\label{career}
\end{figure}

The availability of digitised text, however, enables also the observation of the textual extension of the Matthew effect in citation rates, or alternatively, the large-scale ``semantic'' version of the Matthew effect in science~\cite{perc_sr13}. By using information provided in the titles and abstracts of over half a million publications that were published by the American Physical Society during the past 119 years, and by identifying all unique words and phrases and determining their monthly usage patterns, it is possible to obtain quantifiable insights into the trends of physics discovery from the end of the 19th century to today (the $n$-gram viewer for publications of the American Physical Society is available at \href{http://www.matjazperc.com/aps}{\textcolor{blue}{matjazperc.com/aps}}). The research revealed that the magnitudes of upward and downward trends yield heavy-tailed distributions, and that their emergence is due to the Matthew effect. This indicates that both the rise and fall of scientific paradigms is driven by robust principles of self-organization, which over time yield large differences in the impact particular discoveries have on subsequent progress. Similar research has also been conducted by Pfeiffer and Hoffmann~\cite{pfeiffer_pnas07}, who analysed the temporal patterns of genes in scientific publications hosted by PubMed. They have observed that researchers predominantly publish on genes that already appeared in many publications. This might be a rewarding strategy for researchers, because there is an obvious positive correlation between the frequency of a gene in scientific publications and the impact of these publications~\cite{pfeiffer_pnas07}. In a way, the Matthew effect can thus be engineered, or at least facilitated, by focusing on the ``hot topics'' in a specific field of research.

Figure~\ref{maps} reveals that the Matthew effect in the impact of scientific research translates also to geography~\cite{perc_sr13}, where the U.S. and large contingents of Europe were able to set the pace in the production of physics research over extended periods of time, interrupted only by periods of war. The collapse of the Soviet Union, the fall of the Berlin Wall, and the related changes in World order during the 1980s and 90s, however, contributed significantly to the globalisation, so that today countries like China, Russia, South America and Australia all contribute markedly to the production of physics. However, a beautiful citation map of the world produced by Pan et al.~\cite{pan_sr12}, where the area of each country is scaled and deformed according to the number of citations received, still reveals a strongly biased geographical distribution of impact. Notably, an in-depth analysis of the scientific production and consumption of physics revealed that even cities can be pinpointed based on their leading positions for scholarly research~\cite{zhang_sr13}. Although for now research along this line seems to be focused predominantly on physics, the applied methodology certainly opens up the possibility for comparative studies across different disciplines and research areas, where the Matthew effect is still to be either confirmed or refuted.

\section{Career longevity}
\label{longevity}

The overwhelming evidence in favour of the Matthew effect in science, affecting the patterns of collaboration, the propagation of citations, and ultimately also scientific progress and impact, probably make it little surprising that the same effect affects also career longevity. Importantly, not just the longevity of scientific careers, but also the longevity of careers in professional sport, as demonstrated in~\cite{petersen_pnas11}.

Career longevity is a fundamental metric that influences the overall legacy of an employee, since for most individuals the measure of success is closely related to the length of their career. In particular, the more successful an individual, the longer his or her career is going to last. Using this as motivation, Petersen et al.~\cite{petersen_pnas11} have analysed publication careers within six high-impact journals, including \textit{Nature}, \textit{Science}, \textit{Proceedings of the National Academy of Sciences}, \textit{Physical Review Letters}, \textit{New England
Journal of Medicine}, and \textit{Cell}, as well as sports careers within four different leagues, including Major League Baseball, Korean Professional Baseball, the National Basketball
Association, and the English Premier League. The conducted research delivered testable evidence in favour of the Matthew effect, wherein the longevity and past success of an individual lead to a cumulative advantage in further developing his or her career~\cite{petersen_pnas11}. From the methodological point of view, it is worth pointing out that for science and professional sports there exist well-defined metrics that quantify career longevity, success, and prowess, which together enable a relatively clear and unbiased assessment of the overall success of each individual employee. In many other professions, however, these criteria are significantly more vague, and thus the same research agenda could be difficult to execute.

To support their quantitative demonstration of the Matthew effect in career longevity, Petersen et al.~\cite{petersen_pnas11} have also developed an exactly solvable stochastic career progress model, which is schematically illustrated and summarized in Fig.~\ref{career}. Model predictions have been validated on the careers of $400\thinspace000$ scientists and $20\thinspace000$ professional athletes. The authors have emphasized the importance of early career development, showing that many careers are stunted by the relative disadvantage associated with inexperience. This is closely related to the workings of the Matthew effect in education (see Section~\ref{educate}), where tests suggest that falling behind in literacy during formative primary school years creates disadvantages that may be difficult to compensate all the way to adulthood~\cite{stanovich_je08}.

\section{Common words and phrases}
\label{words}
Moving away from scientific production and impact for good, in this section we review recent research related to the evolution of the most common English words and phrases~\cite{perc_jrsi12}. Already during the 60s, the economist Herbert Simon and the mathematician Beno{\^i}t Mandelbrot had a dispute over the origin of the power-law distribution of word frequencies in text~\cite{zipf_49, tsonis_c97, ferrer_jql01, cancho_epjb05, bernhardsson_njp09, serrano_pone09}. Simon defended the role of randomness and preferential attachment, while Mandelbrot argued in favour of an optimisation framework~\cite{kornai_08}. The original proposal made by Zipf, on the other hand, was that there is tension between the efforts of the speaker and the listener, and it has been shown by means of mathematical modelling that this may indeed explain the origins of scaling in the usage of words \cite{cancho_pnas03}. The ecophysics of language change \cite{sole_jrsi10} --- the application of models from statistical physics and theoretical ecology to the study of language dynamics --- has since evolved into a beautiful and vibrant avenue of research~\cite{sole_n05, baronchelli_jsm06, dall_pre06, loreto_np07, puglisi_pnas08, baronchelli_pone12, baronchelli_ac12, mocanu_pone13}.

\begin{figure}
\centering{\includegraphics[width = 8.5cm]{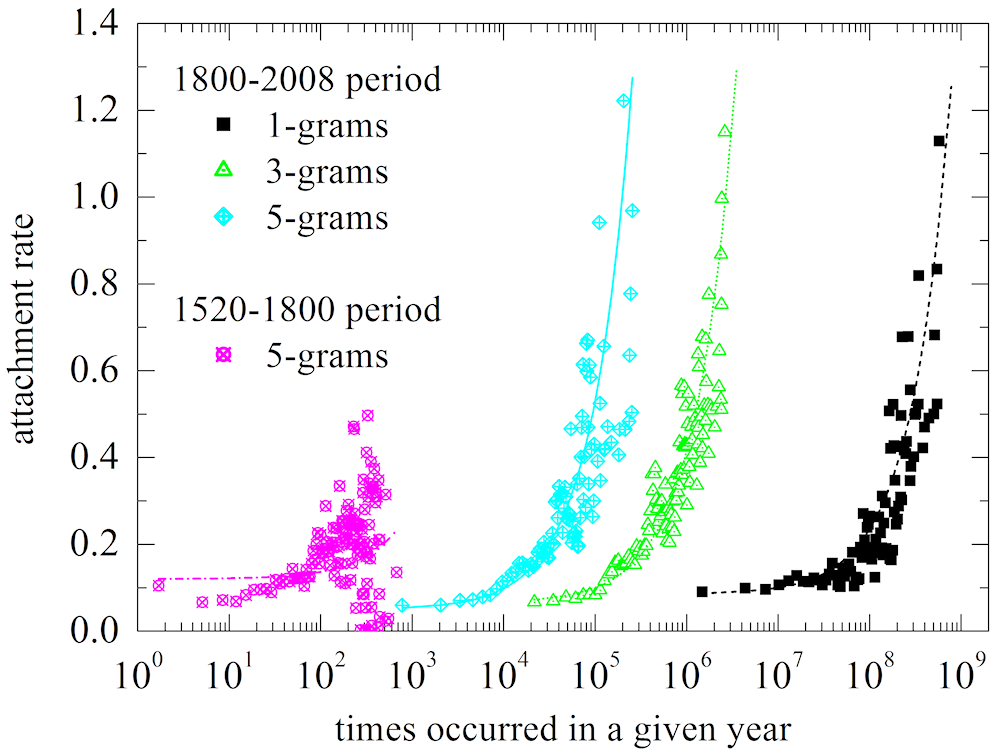}}
\caption{Emergence of linear preferential attachment in the evolution of the most common English words and phrases during the past two centuries. Two time periods were considered separately, as indicated in the figure legend. While preferential attachment appears to have been in place already during the $1520-1800$ period, large deviations from the linear dependence (the goodness-of-fit is $\approx 0.05$) hint towards inconsistencies that may have resulted in heavily fluctuated rankings. The same analysis for the 19th and the 20th century provides much more conclusive results. For all $n$ the data fall nicely onto straight lines (the goodness-of-fit is $\approx 0.8$), thus indicating that the Matthew effect might have shaped the large-scale organization of the writing of English books over the past two centuries. The figure is reproduced from~\cite{perc_jrsi12}.}
\label{attach}
\end{figure}

A direct test for preferential attachment in the evolution of the most common English words and phrases~\cite{perc_jrsi12} was made possible by the work of Michel et al.~\cite{michel_s11}, which was accompanied by the release of a vast amount of data comprised of metrics derived from $\sim4\%$ of books ever published. Raw data, along with usage instructions, is available and updated at \href{http://books.google.com/ngrams/datasets}{\textcolor{blue}{books.google.com/ngrams/datasets}} as counts of $n$-grams that appeared in various book corpora over the past centuries with a yearly resolution. By recursively scanning all the files from the English corpus in the search for those $n$-grams that had the highest usage frequency in any given year, it is possible to determine the most common English words and phrases with a yearly resolution. Tables listing the top $100$, top $1000$ and top $10\thinspace000$ $n$-grams for all available years since $1520$ inclusive, along with their yearly usage frequencies and direct links to the Google Books Ngram Viewer, are available at \href{http://www.matjazperc.com/ngrams}{\textcolor{blue}{matjazperc.com/ngrams}}. From this, it is possible to derive evidence in favour of preferential attachment as shown in Fig.~\ref{attach}, which indicate that the higher the number of occurrences of any given $n$-gram, the higher the probability that it will occur even more frequently in the future. More precisely, for the past two centuries the points quantifying the attachment rate follow a linear dependence, thus confirming that the Matthew effect is behind the power-law distribution of word frequencies in text, as argued by Herbert Simon. Evidently, this does not rule out an optimisation framework that was favoured by Beno{\^i}t Mandelbrot, as preferential attachment itself might be the outcome of optimisation~\cite{souza_pnas07, papadopoulos_n12, barabasi_n12}.

Somewhat related to the study of the most common English words and phrases is also the study of popular memes, which has recently attracted considerable attention~\cite{adar_ieee05, leskovec_acm09, gomez_kdd10, simmons_aaai11, conover_aaai11, weng_sr12, gleeson_prl14}. According to Dawkins, memes are the cultural equivalent of genes that spread across the human culture by means of imitation \cite{dawkins_89}. The competition among memes has been studied by Weng et al.~\cite{weng_sr12}, who by means of an agent-based model accounting for the dynamics of information diffusion, showed that in a world with limited attention only a few memes go viral while most do not. These predictions are consistent with empirical data from Twitter, and they explain the massive heterogeneity in the popularity and persistence of memes as deriving from a combination of the competition for our limited attention and the structure of the social network, without the need to assume different intrinsic values among ideas~\cite{weng_sr12}. The study of how memes compete with each other for the limited and fluctuating resource of user attention has also amassed the attention of physicists, who showed that the competition between memes can bring a social network at the brink of criticality \cite{stanley_71}, where even minute disturbances can lead to avalanches of events that make a certain meme go viral \cite{gleeson_prl14}.

\section{Education and beyond}
\label{educate}
In addition to the above-reviewed examples of the Matthew effect in empirical data, there exist many more, related for example to education~\cite{stanovich_je08} and brain development~\cite{raizada_fhn10}, which we here review in passing for a more complete coverage of the subject.

In his synthesis titled \textit{Matthew effects in reading: Some consequences of individual differences in the acquisition of literacy}~\cite{stanovich_je08}, Stanovich presents a framework for conceptualising the development of individual differences in reading ability, with special emphasis on the concepts of reciprocal relationships --- situations where the causal connection between reading ability and the efficiency of a cognitive process is bidirectional, and on organism-environment correlation --- the fact that differentially advantaged organisms are exposed to non-random distributions of environmental quality. Foremost, it is explained how these mechanisms operate to create the rich-get-richer and the poor-get-poorer patterns of reading achievement, and the framework is used to explicate some persisting problems in the literature on reading disability and to conceptualise remediation efforts in reading. Due to the Matthew effect, early deficiencies in literacy may bread lifelong problems in learning new skills, and falling behind during formative primary school years may create disadvantages that could be difficult to compensate all the way to adulthood~\cite{stanovich_je08}. It must be noted, however, that the degree to which the Matthew effect actually holds true in reading development is a topic of considerable debate~\cite{shaywitz_m95, scarborough_m03, morgan_m08}.

The review by Raizada and Kishiyama on the effects of socioeconomic status on brain development~\cite{raizada_fhn10} also draws on the Matthew effect, in particular as a potential triggering mechanism for a long-term self-reinforcing trend in training executive function in young children, with improved self-control enabling greater attentiveness and learning, which would in turn help to make a child's educational experiences more rewarding, thereby facilitating yet more intellectual growth. The authors are sceptical about this rather ``rosy-sounding'' scenario, but note that specific interventions aimed at improving the cognitive development of children with low socioeconomic status may well trigger the desired effect. Indeed, Cohen and colleagues have shown that even brief self-affirmation writing assignments aimed at reducing feelings of academic threat in ethnic minority high-school students had the effect of producing significant improvements in grade-point average, which endured over a period of $2$ years \cite{cohen_s06, cohen_s09} --- a potential indication that the Matthew effect might have kicked in.

As noted in the Introduction, the concept today is in wide use to describe the general pattern of self-reinforcing inequality that can be related to economic wealth, political power, prestige and stardom. Although these examples are to a degree rooted in folktales and lack firm quantitative support, they can nevertheless be supported by plausible arguments in favour of the Matthew effect. Being born into poverty, for example, greatly increases the probability of remaining poor, and each further disadvantage makes it increasingly difficult to escape the economic undertow. The Matthew effect also contributes to a number of other concepts in the social sciences that may be broadly characterized as social spirals. Economists speak of inflationary spirals, spiralling unemployment, and spiralling debt. These spirals exemplify positive feedback loops, in which processes feed upon themselves in such a way as to cause nonlinear patterns of growth. To make a complete account of such examples exceeds the scope of this review, and so we are content to draw from the recent book \textit{The Matthew effect: How advantage begets further advantage} by Rigney~\cite{rigney_13}, which we warmly recommend to interested readers.

\section{Discussion}
\label{discuss}
As we hope this review shows, the Matthew effect is puzzling yet ubiquitous across social and natural sciences. It affects patterns of scientific collaboration, the growth of socio-technical and biological networks, the propagation of citations, scientific progress and impact, career longevity, the evolution of the most common words and phrases, education, as well as many other aspects of human culture. The recently acquired prominence of the Matthew effect is largely due to the rise of network science~\cite{barabasi_np12}, and the concept of preferential attachment in particular~\cite{barabasi_s99}. Accordingly, the title of this review might as well have been ``Preferential attachment in empirical data'', but since the Matthew effect describes more loosely the general principle that advantage tends to beget further advantage, the age-old Matthew ``rich-get-richer'' effect ultimately won the toss.

The theory of evolving networks based on growth and preferential attachment was motivated by extensive empirical evidence documenting the scale-free nature of the degree distribution, from the cell to the World Wide Web, and it was this theory, along with the ever increasing availability of digitised data at the turn of the 21st century, that ultimately led to the development of the methodology for measuring preferential attachment and the subsequent application of these methods on a wide variety of complex systems. Although the progress made during the past decade related to data-based mathematical models of complex systems has been truly remarkable, the data explosion we witness today is surely going to accelerate research along this line even more. Indeed, ``big data''~\cite{rajaraman_12} is the keyword for current complex systems research, and the data windfall is also surely going to promote research on the Matthew effect. Especially data from social media, but also from neuroscience as well as electronic publication and informatics archives, offer many opportunities for fascinating scientific discoveries in the nearest future.

Concepts such as preferential attachment, cumulative advantage, and the Matthew effect are at the heart of self-organization in biology and societies, and they give rise to emergent properties that are impossible to understand, let alone predict, at the level of constituent agents. The emergent collective modes of behaviour are due to the heterogeneity of the interaction patterns, the presence of nonlinearity and feedback effects, and it is here were the reasons behind the Matthew effect ought to be sought. This, however, raises the question whether the Matthew effect is due to chance or optimisation~\cite{souza_pnas07, papadopoulos_n12, barabasi_n12}. While theoretical models in general rely or dumb luck to yield the power laws, in reviewing the subject on empirical data, one finds it difficult to believe that the selection of a collaborator or a sexual partner, or the hiring for a tenure-track position, would be left to chance. These decisions certainly do depend also on unpredictable factors, but predominantly they are nevertheless based on factors such as common appeal, competence, and prowess. The argument in favour of randomness gains traction when cognition and reasoning obviously no longer apply --- consider the emergence of hubs in protein-interaction networks through gene duplication \cite{pastor_jtb03} (see also~\cite{eisenberg_prl03} for a thorough discussion). But more often than not, the line between chance and thought is much more blurred, like by the propagation of citations. Common sense tells us that credit should be given where credit is due, yet researchers often cite a paper just because it has been cited many times before. An interesting discussion of this was recently delivered by Golosovsky and Solomon~\cite{golosovsky_jsp13}, who concluded that such spreading of citations and ideas is akin to the epidemiological process \cite{bettencourt_pa06} and to the copying mechanism  \cite{krapivsky_pre05}. Google Scholar has even been criticized for strengthening the Matthew effect by putting high weight on citation counts in its ranking algorithm~\cite{beel_g09}, by means of which highly cited papers that appear in top positions gain ever more citations while new papers hardly appear in top positions and therefore struggle to amass new citations. Ultimately, one ends up agreeing with Barab{\'a}si~\cite{barabasi_n12}, who noted that we do not need to choose between luck and reason in preferential attachment, but simply strive towards a deeper understanding of this puzzling yet ubiquitous force.

The Matthew effect is obviously at the interface of many different fields of research, and while its potential has been realized in the realm of complex systems as being one in a series of fundamental laws that determine and limit their behaviour, the concept deserves also to reach a wider audience and to inform public policy decisions that have an impact on inequality in areas such as taxation, civil rights and public goods~\cite{pacheco_c14, perc_jrsi13}.

\begin{acknowledgments}
This work was supported by the Slovenian Research Agency (Grant P5-0027). Special thanks go to Esteban Moro for his insightful tutorial that enabled the creation of Fig.~\ref{esteban} and the pertaining videos.
\end{acknowledgments}

\end{document}